# Fireballs in Dense Stellar Regions as an Explanation of Gamma Ray Bursts

Nir J. Shaviv & Arnon Dar
*Department of Physics, Israel Institute of Technology, Haifa 32000, Israel*

**ABSTRACT**
We study a cosmological scenario for gamma ray bursts (GRBs) where relativistic flows interact with dense radiation fields. It is shown that this scenario is plausible in very dense stellar regions which are known to exist in collapsed cores of globular clusters or dense nuclei of galaxies. It yields a correct quantitative description of the temporal behavior of GRBs. Several other properties of GRBs are easily explained.

## 1 INTRODUCTION

Since their discovery, Gamma-Ray Bursts (GRBs) have remained an unsolved puzzle. Many of their features still pose difficult problems. If originating at cosmological distances, as might be suggested by their remarkable isotropy (Meegan et al. 1992), then their production involves an enormous release of energy in a short time. Their short durations imply very compact sources and enormous pair production, which makes them optically thick to their own $\gamma$-ray emission. Hence, the original $\gamma$-rays are concealed by an opaque envelope, which reduces the original $\gamma$-rays to much softer thermal photons (Cavallo & Rees 1978, Paczynski 1986, Goodman 1986, and Goodman, Dar & Nussinov 1987). They are emitted when eventually the fireball expands enough for the optical depth to decrease below unity. Although the expansion does not necessarily increase the duration because of its Lorentz beaming of the emitted radiation, it cannot explain the non-thermal spectrum of GRBs. Furthermore, baryonic contamination greatly affects its dynamics. Even a baryon load as small as $M_{\mathrm{baryon}} \sim 10^{-9} M_{\odot}$ can convert most of the energy stored in the radiation to kinetic energy of the baryons (Paczynski 1990, Shemi & Piran 1990).

It turns out, that both the non-thermal spectrum and the effects of the baryon overload can still agree with fireball scenarios if the kinetic energy of the baryons is converted back into radiation. Several such mechanisms exist. For instance, Rees & Mészáros (1992) suggest the deceleration of the relativistic wind by the ambient interstellar matter, which is compressed in relativistic shock waves and then emits $\gamma$-rays while cooling. An alternative mechanism for reconversion of the kinetic energy into $\gamma$-rays was proposed by Shemi (1993a) where the interaction of the relativistic wind is with a local thermal radiation field. The relativistic electrons up scatter the ambient photons, and shift them towards the direction of flow, modifying their spectrum into a power law. The rather large scale of $\lesssim 10^7$ light-seconds is still transformed into a short burst because of the relativistic beaming. Furthermore, if the wind is not in a form of a fireball, but of a jet, then the large spectrum of durations can easily be explained as a geometrical effect (Shaviv & Dar 1995). In his work, Shemi (1994) has shown that Compton drag of relativistic debris of cosmic fireballs can explain several generic problems in optically thick GRBs - their compactness, non-thermal spectrum, effects of baryonic loads, and time scales. It has though one very major drawback - it needs very dense radiation fields, those occurring only in dense cores of globular cluster and in galactic nuclei.

We first consider the scenario proposed by Shemi (1993a, 1994) of a thin fireball shell (as is shown to form by Shemi 1993b, Piran, Shemi & Narayan 1993 and Mészáros, Laguna & Rees 1993) scattering photons of the radiation field in a dense stellar cluster. We then show that in addition to explaining various properties of GRBs, it yields time profiles which agree remarkably well with the experimental observations.

## 2 ENERGETICS OF RADIATIONALLY DECELERATED FIREBALLS

How plausible is it to find high photon density regions? One has to go only a few kpc's to reach the collapsed cores of several globular clusters (about 20% of the globular clusters are core collapsed, see e.g. Chernoff & Djorgovski 1989) or the centre of our galaxy. In several globular clusters which had their core collapsed, one can find within a sphere of radius 0.1-pc a large fraction of the cluster's mass, giving very high stellar densities (exceeding even $10^7 M_{\odot}\mathrm{pc}^{-3}$, e.g., Phinney & Sigurdsson 1991). Unfortunately, the centre of our galaxy is blocked by gas and dust, but in the galaxy of Andromeda, where direct visible measurements can be made, there exists a region of 5-pc by 10-pc which contains a total mass of $10^7$ to $10^8$ $M_\odot$, corresponding to a stellar density of about $2 \times 10^5 M_\odot \mathrm{pc}^{-3}$. As we shall see, these types of stellar densities can give rise to an ample photon field, capable of transforming a great fraction of the kinetic energy of fireballs into $\gamma$-rays.

It is interesting to note that the dynamics of collapsed cores of globular clusters must give rise to hard binaries which stop the unstable collapse (e.g. Heggie 1980). These abundant binaries are the probable cause for the relatively very large number of 'interesting' objects such as low mass X-ray binaries and binary neutron stars. Among the four known systems of neutron star binaries, PSR 2127+11C is in the globular cluster M15 (Prince et al. 1991), in contrast with the total mass in globulars which is only $10^{-3}$ of the total galactic mass. It is thus very plausible that these dense clusters hold within them many GRB progenitors, much more than can be estimated from their mass fraction [1].

If we look at such a dense cluster having a stellar population of $N_\star$ within a region of radius $R = 0.1\ d_{0.1}$pc ($d_{0.1}$ measures the radius in units of 0.1-pc), and an average stellar luminosity $\ell L_\odot$ with photons of energy $\varepsilon_0$-eV, then the photon density at the centre is:

$$n_\gamma = \frac{3 N_\star \ell L_\odot}{4\pi c R^2 \varepsilon_0} = 2.0 \times 10^4 \frac{\ell N_{\star,5}}{d_{0.1}^2 \varepsilon_0}\ \gamma\ \mathrm{cm}^{-3}, \quad (1)$$

with $N_\star \equiv 10^5\ N_{\star,5}$. If we assume such a photon density, and that the only means of interaction is by Compton scattering, then we can calculate the amount of energy lost by an electron while traversing a path of length $R$ (i.e. the average distance until it leaves the cluster).

The electron's energy is: $E_e = \gamma m_e c^2$. On the average, an electron loses $E_\gamma = \frac{4}{3}\gamma^2 \epsilon_0$ of its energy per collision. The mean free path is $\lambda = 1/\sigma_\mathrm{T} n_\gamma$ where $\sigma_\mathrm{T} = 0.66 \times 10^{-24}\mathrm{cm}^2$ is the Thomson cross section, and the average number of interactions along $R$ is $R/\lambda = R\sigma_\mathrm{T} n_\gamma$. Therefore, the electron will lose an average energy of $\frac{4}{3}\gamma^2 \epsilon_0 R \sigma_\mathrm{T} n_\gamma$ before leaving the cluster. Its fractional energy loss, $\zeta$, is given by:

$$\zeta = \frac{4\sigma_\mathrm{T} n_\gamma R \gamma \epsilon_0}{3 m_\mathrm{eff} c^2}, \quad (2)$$

with $m_\mathrm{eff}$ being the effective mass associated with each electron. For a baryonic fireball, $m_\mathrm{eff} = m_e + m_p \approx m_p$, but if the fireball is still rich in pairs, the effective mass can be lower, reaching $m_e$ for $n_\mathrm{pairs} >> 1000\ n_p$.

Since the duration of the GRB is given by $t_0 \approx \gamma^{-2}(R/2c)$, one obtains:

$$\zeta = 4.0 \times 10^{-8} \ell N_{\star,5} t_0^{-1/2} d_{0.1}^{-1/2} \mu^{-1}, \quad (3)$$

with $\mu \equiv m_\mathrm{eff}/m_p$. For the Andromeda galaxy where $N_{\star,5} \approx 1000$, $d_{0.1}^{-1/2} \approx 0.1$ and $t_0^{-1/2} = 0.3$ or 3 for 10 sec or 0.1 sec bursts, eq. (3) for $\mu = 1$ yields $\zeta = 0.1 \times 10^{-5}$ or $1 \times 10^{-5}$, respectively. If the observed $\gamma$-ray energy is $E_{50} \times 10^{50}$-erg then the initial kinetic energy of the fireball, assuming *no* beaming, is $E_\mathrm{kin} = 50 E_{50}\ M_\odot c^2$ or $5 E_{50}\ M_\odot c^2$, respectively. For collapsed cored globular clusters (such as M15) we have smaller values: $N_{\star,5} \approx 5$ and $d_{0.1}^{-1/2} \approx 1$ yielding: $\zeta = 0.5 \times 10^{-7}$ *or* $5 \times 10^{-7}$ for $\mu = 1$, giving $E_\mathrm{kin} = 1000 E_{50}$ or $100 E_{50}\ M_\odot c^2$, respectively. These numbers are obviously too large, rendering this model impossible unless we either utilize beaming, decrease $\mu$ or increase $\sigma$. If we succeed to form a pair-rich fireball, $\mu$ can reach 2000, in which case the required initial kinetic energy can range from $E_\mathrm{kin} = 2.5 \times 10^{-3} E_{50}\ M_\odot c^2$ to $0.5 E_{50}\ M_\odot c^2$, depending on the parameters, without any beaming.

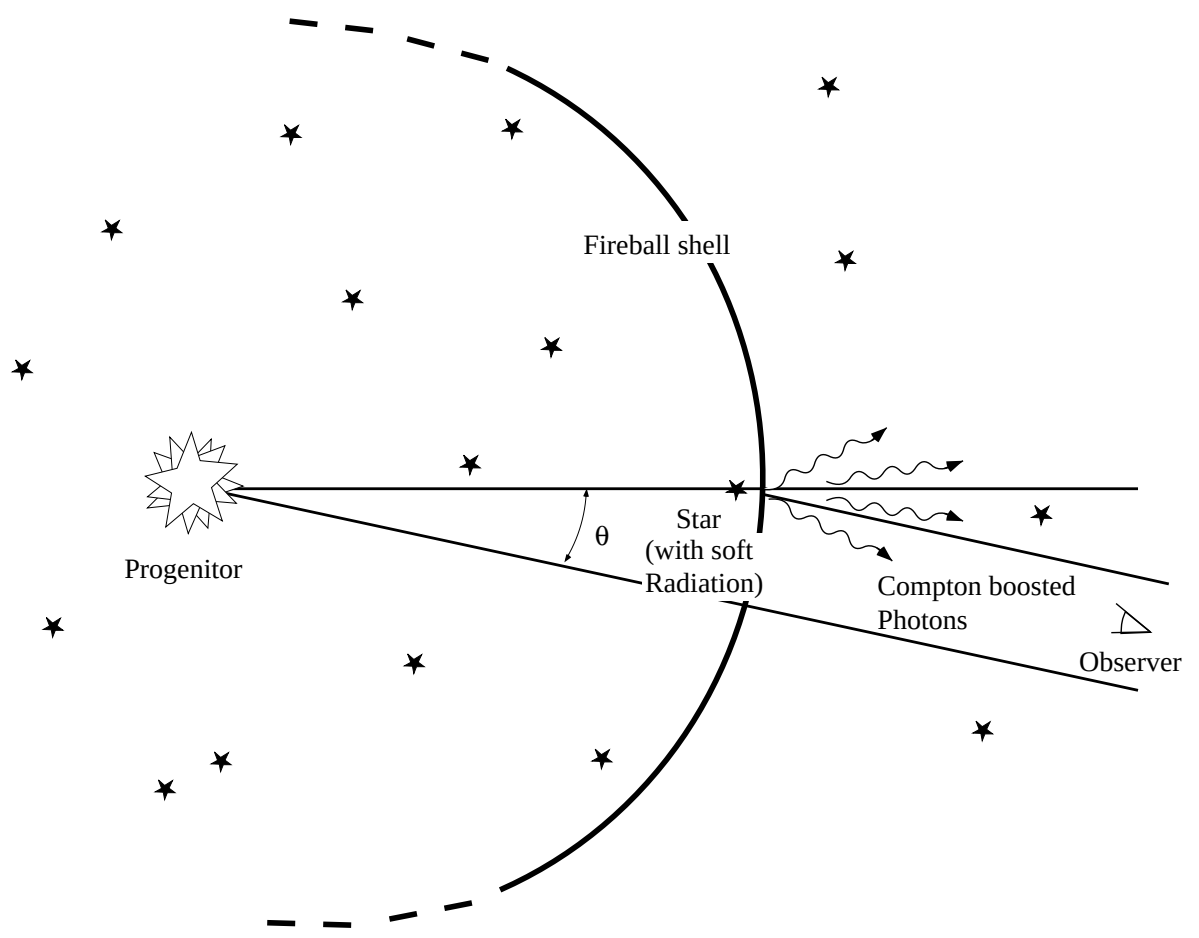

**Figure 1.** A schematic drawing illustrating the proposed scenario. A fireball inside a dense stellar region is formed. While expanding it scatters the cluster's radiation field photons. Most of the observed scattered photons come from the region near the stars, were the light density is the highest.

Having a higher cross-section $\sigma$ for scattering photons off the fireball is a very realistic possibility for NS-NS merger scenarios. It is well known that the crust of neutron stars is composed of mainly $^{56}$Fe or heavier nuclei. A fireball composed of such nuclei will have a very large opacity at a temperature of a few eV [2]. For such nuclei, this opacity (per unit mass in the fireball) formed by the large cross-section for ionization by the ambient photons that in the fireball's rest frame have keV energies, can be greater than $10^5$ times the opacity resulting from Compton Scattering. The ionized levels are quickly filled, emitting back photons that in the observers frame appear as ~ 100-keV photons. Since the different lines are so numerous, the smearing resulting from the different Doppler shifts of the different shells in the fireball will smear out any possible lines. Here, the required initial kinetic energy can range from $E_\mathrm{kin} = 5 \times 10^{-5} E_{50}\ M_\odot c^2$ to $0.01 E_{50}\ M_\odot c^2$. The beaming geometry will reduce the energies even more.

[1] For Instance, the *conservative* estimate by Phinney (1991) of the neutron star merger rate in globular clusters is $3 \times 10^{-9} h^3 \mathrm{Mpc}^{-3}\mathrm{yr}^{-1}$. This rate is smaller by less than an order of magnitude than the rate required to meet the *CGRO–BATSE* event rate

[2] The fireball's temperature can be found by calculating the temperature and radius at which the photon opacity has decreased to unity. Such a calculation (with equations that can be found, for example, in Mészáros, Laguna & Rees 1993) yields a temperature of a few eV. At these temperatures and densities, all atomic levels with energies less than ~ 200-eV will be ionized.

## 3 THE TEMPORAL PROFILE OF A FIREBALL INTERACTING WITH DENSE INTERSTELLAR RADIATION

We shall now proceed to calculate the time profile of a GRB resulting from a fireball shell that substands a large opening angle and expands in a dense stellar field. If we look at Fig 1, we see that two regions can emit most of the observed GRB energy. The first is the region along the observer's line of sight, while the second is the region near a star. The former is a region of angular size $\approx 1/\gamma^2$ with a boosting factor of $\approx \gamma$, while the second is a region of angular size $\theta^2$ with a boosting factor of:

$$\frac{1}{\gamma\theta^2 + 1/\gamma} \approx \frac{1}{\gamma\theta^2} \tag{4}$$

leading to the same result. However, these factors must be weighted with the appropriate distance attenuation, which is larger for the former region. Hence, the received fluence will come mainly from a region near the star. We shall now estimate its time profile. The time dependent intensity profile formed by just one star is given approximately by:

$$\begin{aligned} I_i d\Omega &= \frac{I_{0,i}\sigma_c(\theta)d\Omega}{4\pi}\int_0^{r_{\max}} \frac{2\pi r dr}{(v(t-t_i))^2 + r^2 + r_{\min}^2} \\ &= \frac{I_{0,i}\sigma_c(\theta)d\Omega}{4}\log\left((v(t-t_i)^2 + r_{\min}^2 + r^2)\right)\Big|_0^{r_{\max}} \\ &\approx \frac{I_{0,i}\sigma_c(\theta)d\Omega}{4}\log\left(\frac{(v(t-t_i))^2 + r_{\max}^2}{(v(t-t_i))^2 + r_{\min}^2}\right) \end{aligned} \tag{5}$$

where $\sigma_c(\theta)d\Omega$ is the fraction of radiation scattered into the solid angle $d\Omega$ at an angle of $\theta$ from the observer's line of sight. This cross-section approximately behaves as $\propto 1/(\gamma\theta^2 + 1/\gamma)$. $r_{\min}$ is a lower cutoff that is formed by the finite width of the shell, and it is assumed to be much less than $r_{\max}$. $t_i$ is the time when the shell crosses the star, and $I_{0,i}$ is the intrinsic luminosity of the star. Note that this result is a very rough approximation. Eq. (17) which assumes that different parts of the fireball scatter with the same incident angle $\theta$, essentially approximates the fireball with a round plane. This assumption is justified since most of the scattered radiation comes from a small region of the fireball near the star, which does not see the curvature of the fireball. In addition, it does not take into account the fact that the scattered intensity depends on both the initial spectrum of the photons and the scattering angle. These effects are too complex to be taken into account analytically. Together with the finite width of the fireball they probably tend to deform the resulting time profile.

The Fourier transform of eq. (17) is:

$$\begin{aligned} \tilde{I}_1(\omega) &= \frac{1}{2\pi}\int_{-\infty}^{\infty} I_i(t)e^{i\omega t}dt \\ &= \frac{1}{2\pi}\int_{-\infty}^{\infty}\frac{I_{0,i}\sigma_c(\theta)}{4}\log\left(\frac{(vt)^2 + r_{\max}^2}{(vt)^2 + r_{\min}^2}\right)e^{i\omega(t+t_i)}dt \\ &= \frac{e^{i\omega t_i}}{4\omega}\left(e^{r_{\min}\omega/v} - e^{r_{\max}\omega/v}\right)I_{0,i}\sigma_c(\theta). \end{aligned} \tag{6}$$

It has three distinctive regions of behavior bounded by $\omega_{\min} = v/r_{\max}$ and $\omega_{\max} = v/r_{\min}$. The behavior in each region is given by:

$$\tilde{\mathrm{I}}_1(\omega) = \begin{cases} \frac{I_{0,i}\sigma_c(\theta)r_{\max}}{4v} & \omega << \omega_{\min} \\ \frac{I_{0,i}\sigma_c(\theta)}{4\omega} & \omega_{\min} << \omega << \omega_{\max} \\ \frac{I_{0,i}\sigma_c(\theta)}{4\omega}e^{-r_{\max}\omega/v} & \omega_{\max} << \omega \end{cases} \tag{7}$$

If $r_{\min}$ which is formed from the finite width of the shell is small, then between $\omega_{\min}$ which is of the order of one over the duration of the burst and $\omega_{\max}$, there will be a large region where $\tilde{I}_1$ will have a $1/\omega$ behavior. The total power spectrum in this region is given by:

$$\begin{aligned} P_{total}(\omega) &= \left(\sum_i \tilde{I}_i(\omega)\right)^2 = \sum_i \left|\tilde{I}_i(\omega)\right|^2 \\ &= \left(\sum_i\left(\frac{I_{0,i}\sigma_c(\theta)}{4}\right)^2\right)\frac{1}{\omega^2}. \end{aligned} \tag{8}$$

Since one has to sum different Fourier transforms with random phases one obtains the second equality, which simply states that the power spectrum of a sum of uncorrelated Fourier transforms is just the sum of the different individual power spectra. Note that the total power spectrum in the middle region has a $1/\omega^2$ behavior, which as we shall see later, is the power spectrum behavior of real GRBs. This $1/\omega^2$ behavior will break down for small enough $\omega$'s, smaller than a few times the reciprocal of the duration, and also for large enough $\omega$'s. If the shortest features in GRBs are of $10^{-3}$s, and originate from the finite width of the shell, than the upper frequency at which the law breaks is $\sim 1000$ Hz. A larger shell size will result with a lower upper frequency.

## 4 NUMERICAL SIMULATIONS AND COMPARISON WITH OBSERVATIONS

In addition to the analytical calculations, we have performed several numerical simulations of fireball expansion in dense stellar regions. Although the analytical calculations yield a temporal behavior similar to that observed for GRBs, they contained several approximations which need verification. In addition, the numerical simulations give a general picture of the emerging time profiles, which is usually not evident from just the power law behavior.

In Fig. 2 we present a typical temporal behavior calculated for a Gaussian star cluster with $N_\star = 5\times 10^5$ and with a radius of $d_{0.1} = 1$, $r_{\min} = 10^{-4}d_{0.1}$ and $\gamma = 100$. The temporal profile is normalized to the burst's length of $1/\gamma^2$. The power spectrum of the temporal profile in Fig. 2 is depicted in Fig. 3. Fig. 4 and 5 are the same as Fig. 2 and 3, but with addition of simulated noise. In both power spectra, the $1/\omega^2$ law is evident. In the second case, the power slope can be seen down to frequencies which are lower than the frequency where the signal is below the noise. Figures 6 and 7 show the temporal profile and the power spectrum of an arbitrarily selected *real* GRB taken from the *BATSE* GRB catalogue (Fishman et al. 1994). The simulated profile and the real profile seem to have the same 'qualitative' and 'quantitative' behaviors. That is the case also for other GRBs since the simulated burst's spectral power law behavior is the same as for real GRBs (Shaviv & Dar 1995) - namely a $1/\omega^2$ behavior.

Fig. 8 and 9 show the temporal and spectral result of a Gaussian star cluster with $N_\star = 10^6$ and with a radius of $d_{0.1} = 1$, but with $\gamma = 1000$. Since the fireball seems more squeezed to an outside observer, less stars will pass through it, giving less peaks inside. Note that the simple profile of eq. 17 was used. More realistic assumption on the scattering process and the finite width of the fireball will give more complex asymmetric peaks, as can be seen in the *real* GRB in Fig. 10 (with its Fourier transform in Fig. 11). Actual intensities will depend on the spectral band of the detector and the spectrum of both the scattering electrons and photons, which together with the scattering angle change the outcoming energy of the photons. This effect is probably negligible, since it is a slowly varying function of time, giving a Fourier transform which is 'localized' in frequency space, and therefore does not change the $1/\omega^2$ law when convolved with the previous result.

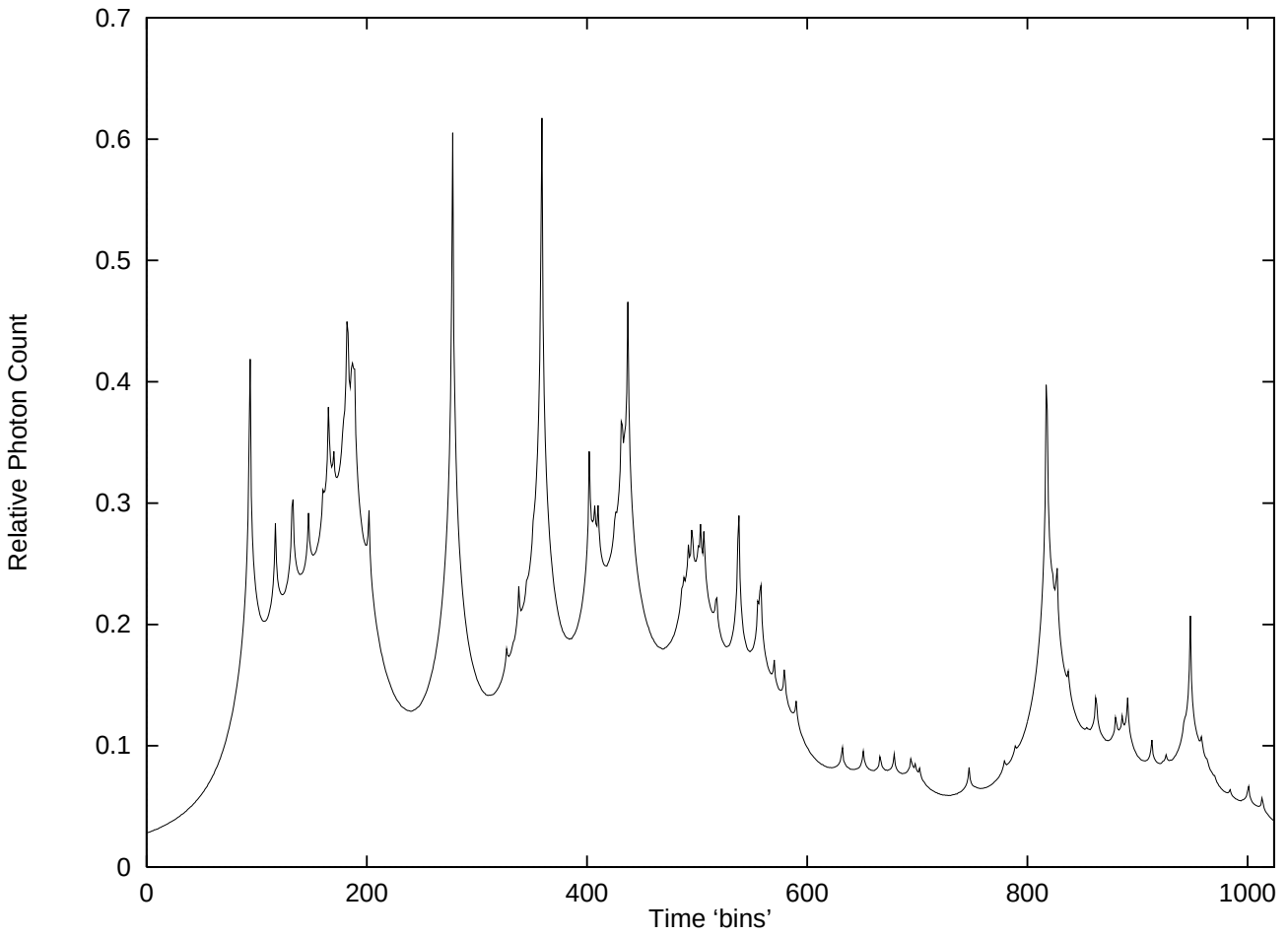

**Figure 2.** The temporal profile of a simulated GRB from a Gaussian star cluster with $N_\star = 5 \times 10^5$ and with a radius of unity. In addition $r_{\min} = 10^{-4}$ and $\gamma = 100$. The temporal profile is normalized to the burst's length of $1/\gamma^2$.

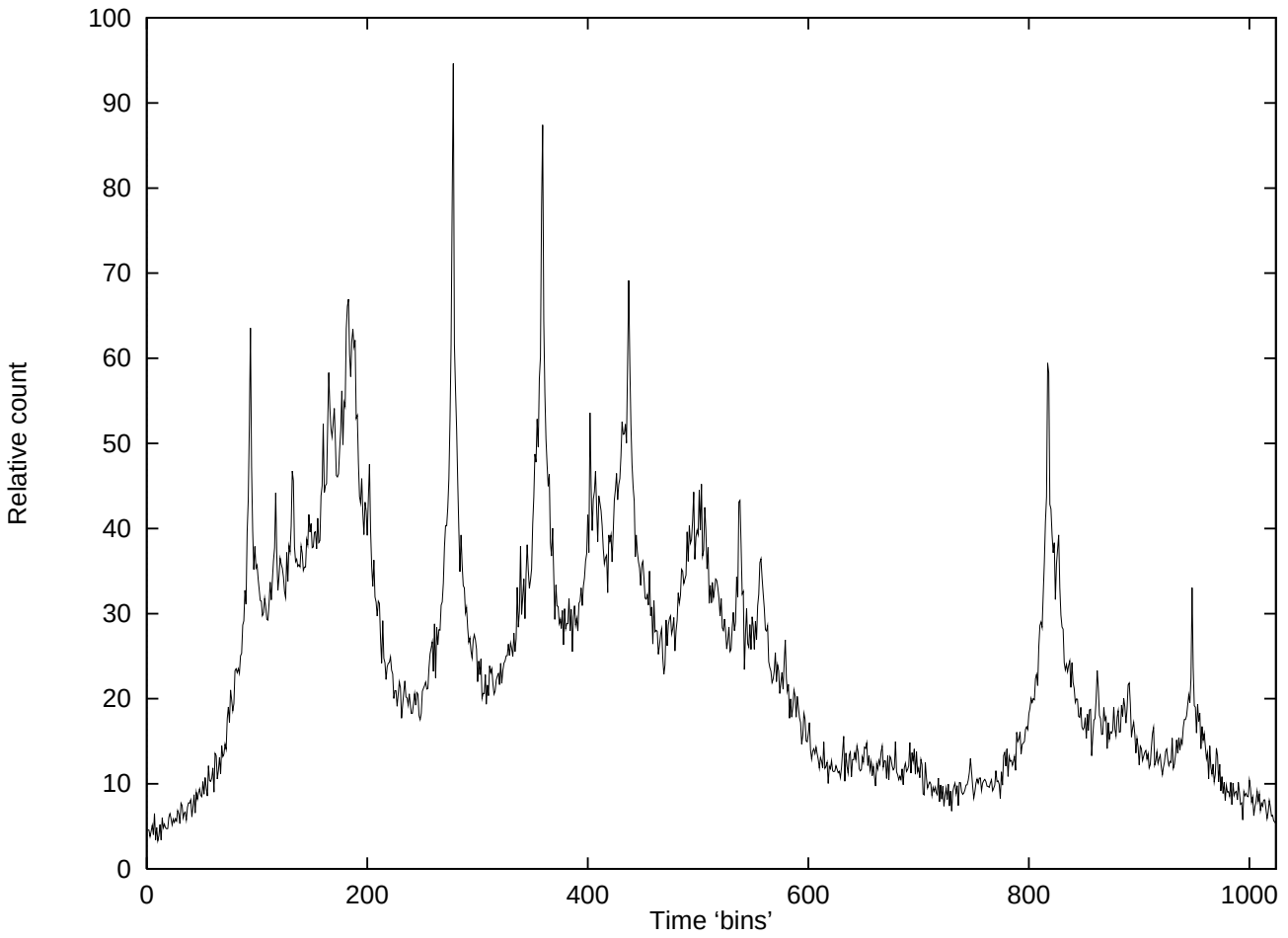

**Figure 4.** Same as Fig. 2, but with a small white noise component added to simulate measured profiles.

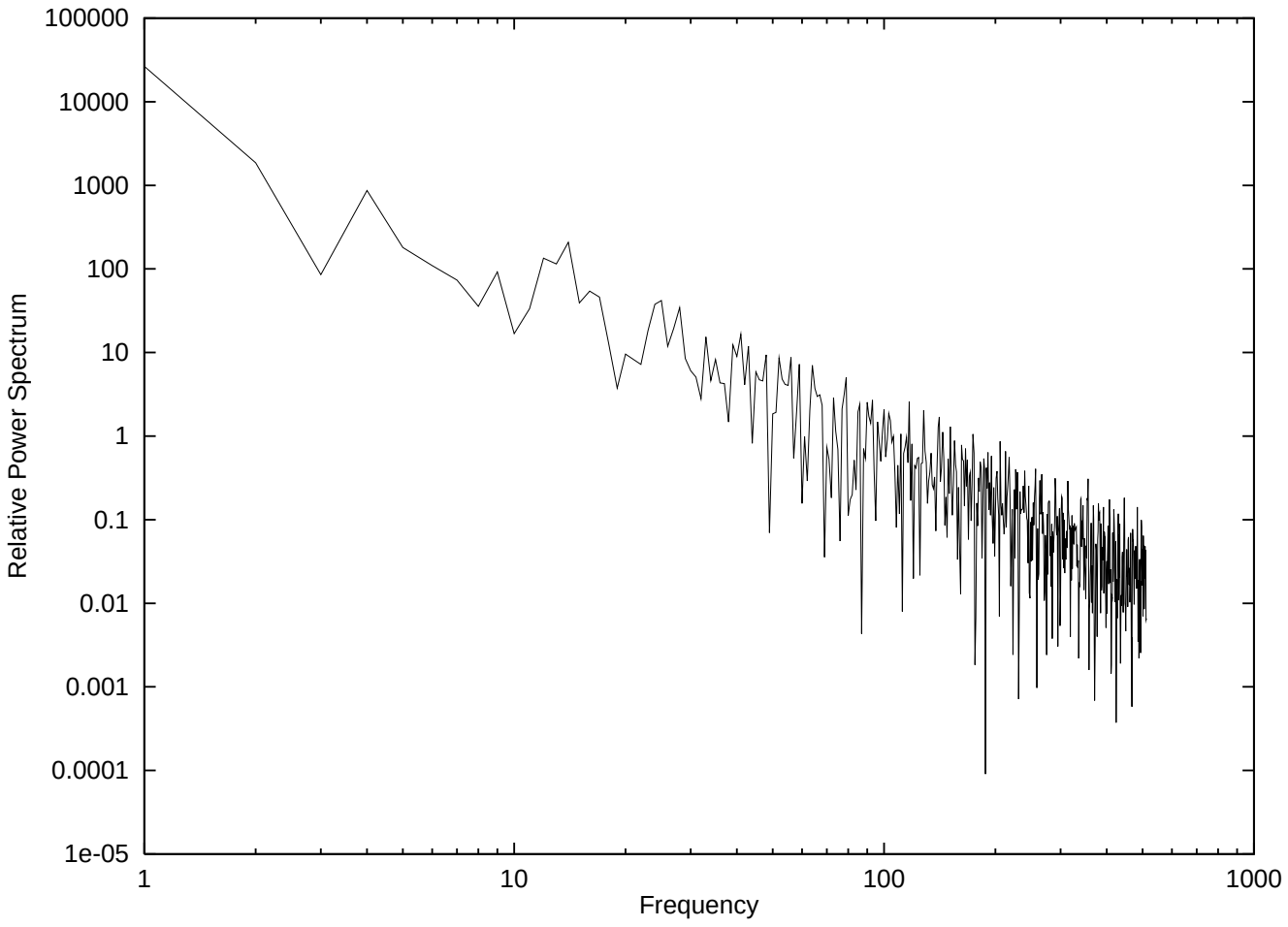

**Figure 3.** The power spectrum of the time profile shown in Fig. 2.

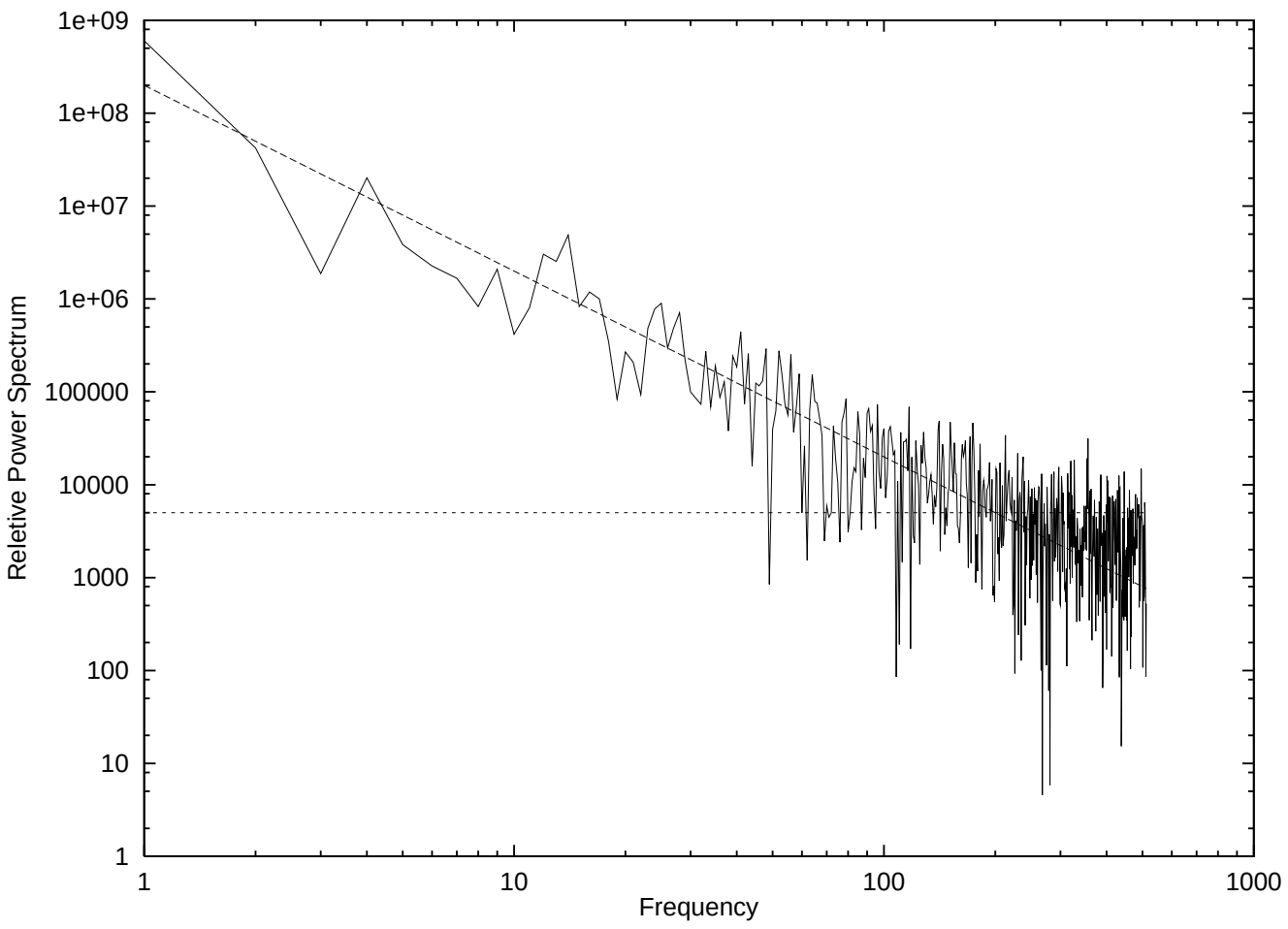

**Figure 5.** The power spectrum of the time profile in Fig. 4. Two lines are drawn, one shows an $\omega^{-2}$ fit, while the other shows the noise level.

The model can explain several additional observations. For instance, there have been several reports of after burst pulses of very energetic photons (e.g. Mukherjee 1994, Hurley et al. 1994). These can come from an X-ray source far from the observer's line of sight, where the Lorentz beaming is too weak to boost the optical radiation into $\gamma$-rays, but can boost the X-rays into hard $\gamma$-rays [3]. Since the source is far from the line of sight, the time for the fireball to reach this X-ray source, as measured by the observer, will be much longer. An additional possibility is the annihilation in flight of positrons left in the fireball. The number density of electrons in the interstellar matter in the galaxy, and certainly in globular clusters, is much lower than the number density of the optical photons in the centre of the cluster. As a consequence, we can expect that only a small fraction of the positrons would annihilate in flight while interacting with electrons of the interstellar medium. These events could also occur out of the dense stellar regions, resulting with delayed hard photons.

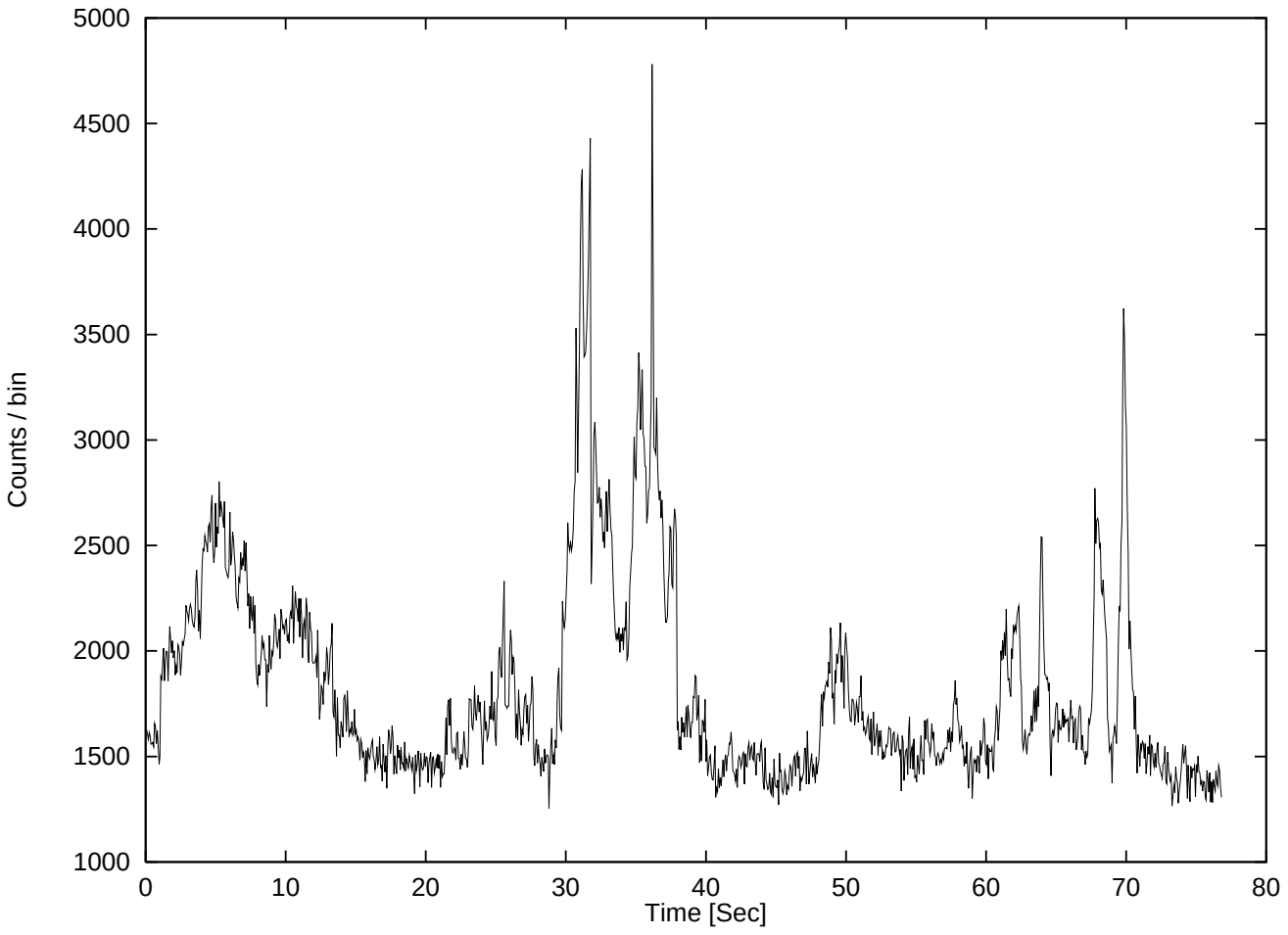

**Figure 6.** The temporal profile of GRB920110 (trigger #1288 of *CGRO–BATSE).*

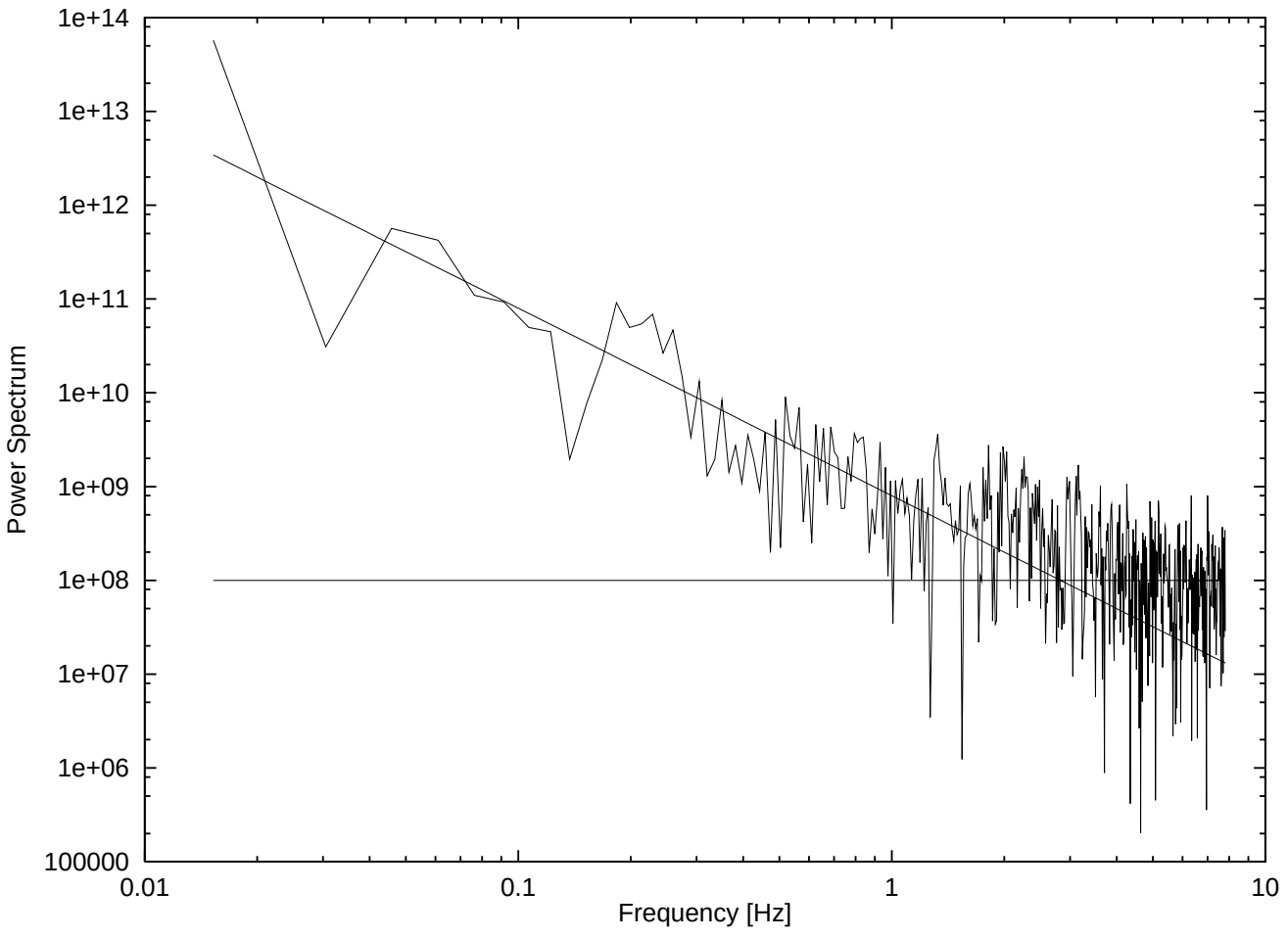

**Figure 7.** The power spectrum of the time profile shown in Fig. 6.

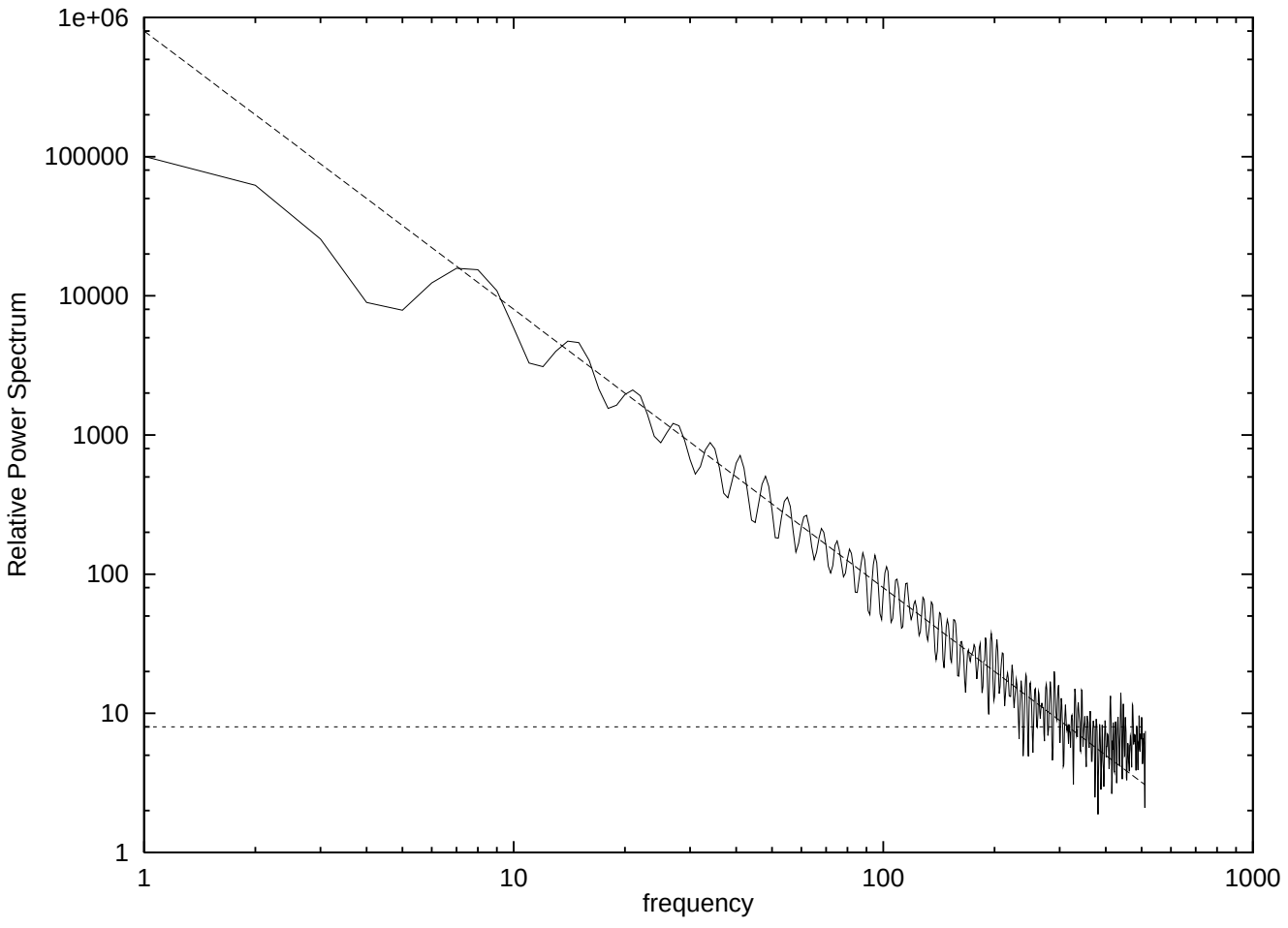

**Figure 9.** The power spectrum of the time profile shown in Fig. 8. Two straight lines are drawn, one is the noise level, while the other has a slope of $\omega^{-2}$.

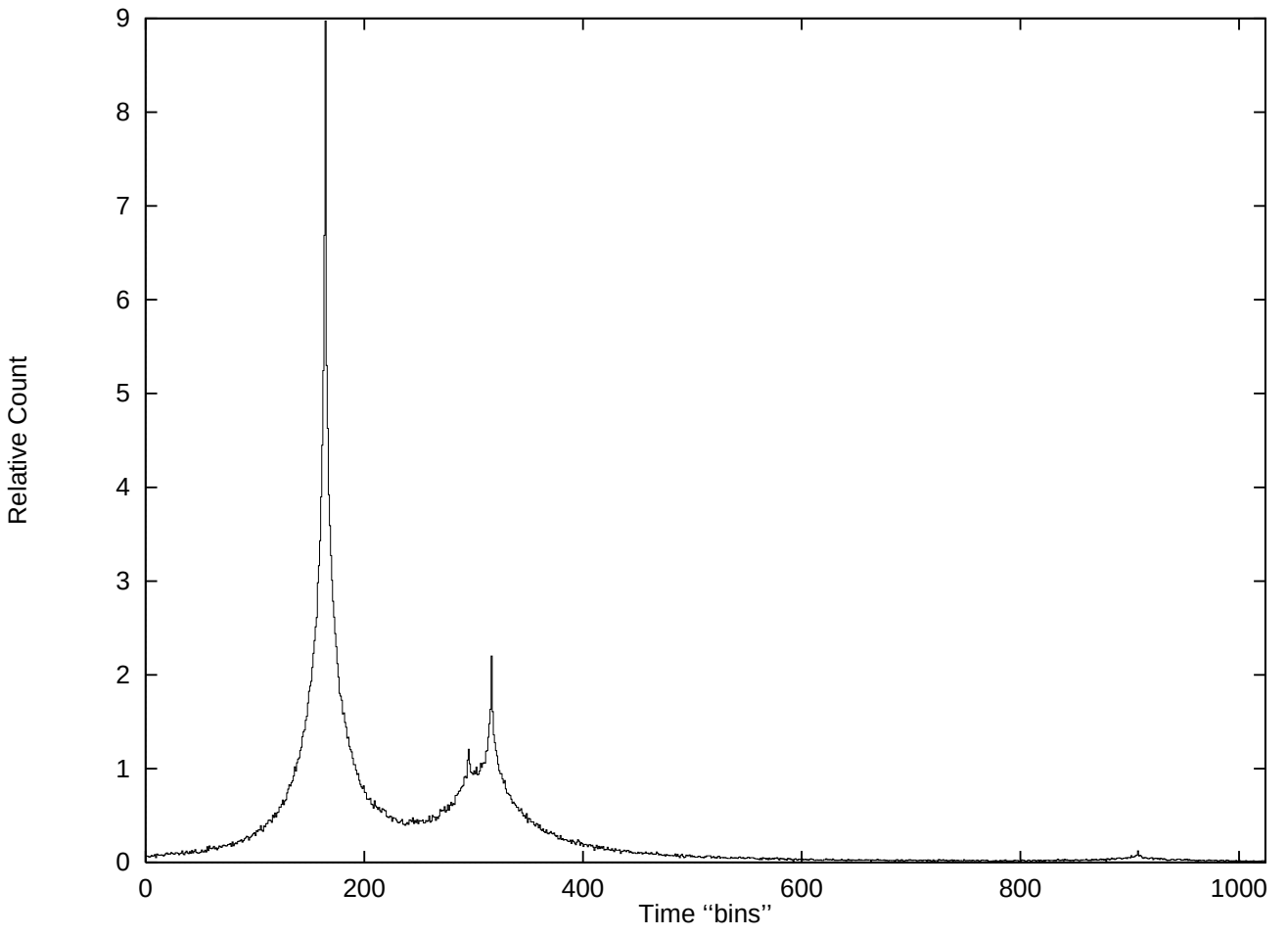

**Figure 8.** The temporal profile of a simulated GRB from a Gaussian star cluster with $N_\star = 10^6$ and with a unit radius. $\gamma = 1000$. The temporal profile is normalized to the burst's duration of $\gamma^{-2}$. A small white noise component was added.

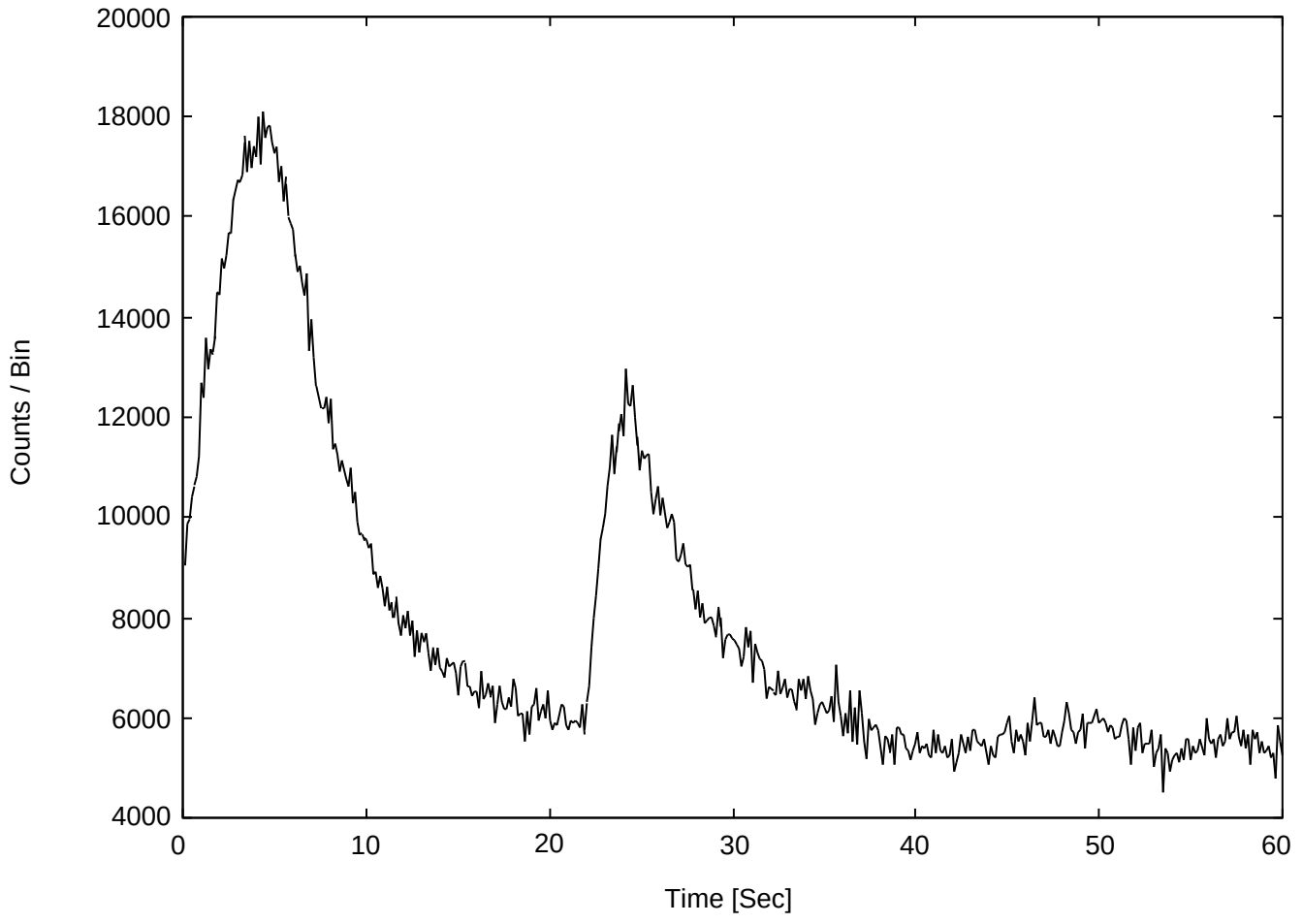

**Figure 10.** The temporal profile of GRB911031 (trigger #973 of *CGRO–BATSE*).

[3] X-ray sources are known to be abundant in globular clusters, see e.g. Katz (1975) and Lewin (1980).

Ford et al. (1995) have reported that there is a general softening of the GRBs with time. Occasionally though, the bursts harden with the increase of the intensity. This also coincides qualitatively with the model. As the wind is decelerated, the lower Lorentz $\gamma$ factor yields on the average a softer spectrum. But, when passing nearby a star, the hardness depends upon the relative incident angle. Occasionally a brighter peak and a harder spectrum will be produced while passing a star.

## 5 DISCUSSION

At first sight, the interaction of a fireball with dense radiation fields seems highly improbable as the source of GRBs. However, we have shown that fireball debris can lose enough of their energy and generate a GRB if occurring in dense stellar fields. Such fields exist in dense galactic nuclei and in the cores of collapsed globular clusters. As was shown, this is possible if we can produce a fireball containing many $e^+e^-$ pairs left after freeze out (as compared with the number of baryons) which do not annihilate. This enables the photons from the radiation field to slow down the massive debris, composed mainly $e^+e^-$ pairs. An additional possibility exists, if the fireball's opacity to the ambient photons is large. This results naturally in fireballs that are composed of high Z material which can originate, for example, from the crust of neutron stars. A problem would occur though, if the interstellar medium is not rarefied. A medium containing too much gas ($n_{\rm ISM} \gtrsim 1$ cm$^{-3}$) can possibly stop the fireball without releasing the energy by boosting the radiation.

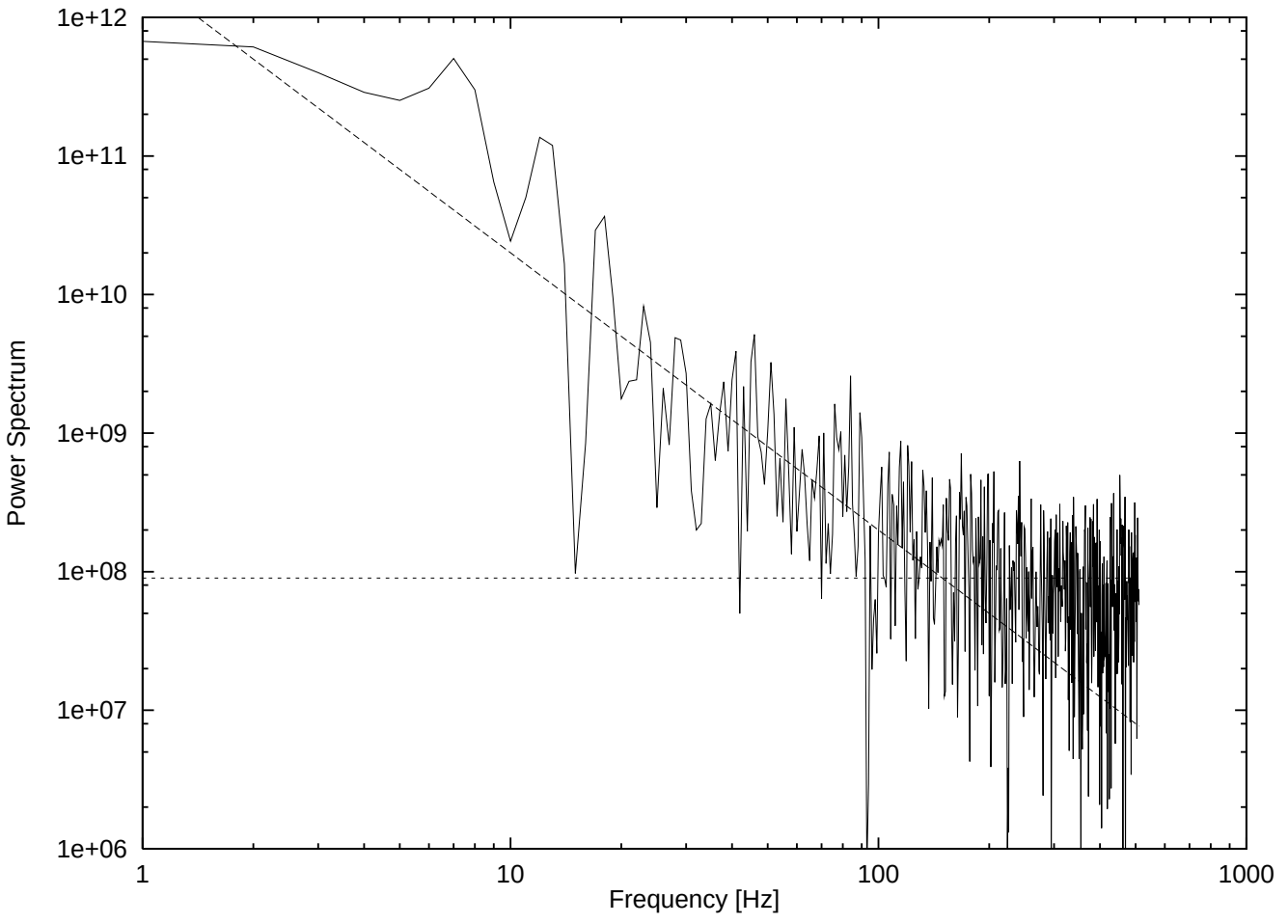

**Figure 11.** The power spectrum of the time profile shown in Fig. 10. The two straight lines are again the noise level and an $\omega^{-2}$ fit.

The structure of the dense radiation fields was taken to be that formed from a random distribution of stars. This was used to calculate the power spectrum of the temporal profile, and gave a power law spectrum with a power index of $-2$ - exactly the same power spectrum of the observed GRBs. It is interesting to note that even if the field contains only a single star, it will produce the same power spectrum; i.e., the power spectrum is not due to the statistical average of many stars, but it is a feature of the individual peeks.

It is clear that the model is still in its infancy, but it already yields important verified predictions. Yet, there are still various GRBs features that have to be explained by the model. One such feature is the spectra of GRBs, which in the model, depends on the average spectrum of the scattered radiation, and the distribution of Lerentz boosting factors in the fireball.

## ACKNOWLEDGMENTS

The authors thank A. Shemi and P. Mészáros for their helpful comments. This research has made use of data obtained through the Compton Observatory Science Support GOF account, provided by the NASA-Goddard Space Flight Center.